# Coexistence of superconductivity and sliding polar metal state in HgPSe$_3$

Xiaohui Yu[1,2,3,8], Wei Zhong[4,8], Saori Kawaguchi[5], Hirokazu Kadobayashi[5], Xiaolin Wang[6], Zhenxiang Cheng[6], Changfeng Chen[7], Binbin Yue[4✉], Jian-Tao Wang[1,2,3✉], Ho-Kwang Mao[4], Fang Hong[1,2,3✉]

[1]*Beijing National Laboratory for Condensed Matter Physics, Institute of Physics, Chinese Academy of Sciences, Beijing 100190, China*

[2]*School of Physical Sciences, University of Chinese Academy of Sciences, Beijing 100190, China*

[3]*Songshan Lake Materials Laboratory, Dongguan, Guangdong 523808, China*

[4]*Center for High Pressure Science & Technology Advanced Research, 10 East Xibeiwang Road, Haidian, Beijing 100094, China*

[5]*SPring-8/JASRI, 1-1-1 Kouto, Sayo-gun, Sayo-cho, Hyogo 679-5198, Japan*

[6]*Institute for Superconducting and Electronic Materials, University of Wollongong, Innovation Campus, Squires Way, North Wollongong, NSW 2500, Australia*

[7]*Department of Physics and Astronomy, University of Nevada, Las Vegas, Nevada 89154, USA*

[8]*These authors contributed equally: Xiaohui Yu, Wei Zhong*

✉Email: yuebb@hpstar.ac.cn; wjt@aphy.iphy.ac.cn; hongfang@iphy.ac.cn

**The simultaneous presence of polarity and metallicity in a material signifies an exotic polar metal state[1], but such materials are extremely rare, especially in bulk form, due to mutually exclusive nature of the fundamental defining properties[2,3]. Here, we report experimental findings that HgPSe$_3$ is a robust bulk polar metal at room temperature with a chiral structure stabilized by pressure and, remarkably, this polar metal hosts superconductivity with critical temperature T$_c$ up to 11 K. Theoretical analysis reveals a two-step interlayer sliding-then-compressing mechanism for coexistence of polarity and metallicity in HgPSe$_3$. This work unveils a new paradigm for creating the bulk polar metal state and simultaneous presence of coexisting quantum orders, raising the prospect of discovering novel emergent physics using pressure as a tuning knob.**

The interplay of competing quantum orders often leads to groundbreaking discoveries of novel physics phenomena, such as multiferroics with ferromagnetism and ferroelectricity[4,5], functional materials with transparency and conductivity[6], novel superconductivity in (anti)ferromagnetic systems[7-9], nontrivial topology[10], abnormal metallic state with low thermal conductivity[11]. Such materials hold great promise and significance to designing next-generation multifunctional devices. Extremely rare among these novel materials are those concurrently hosting polar and metallic states, because the screening of the long-range Coulomb interactions by conduction electrons in metals is highly unfavorable, to the extent of almost mutually exclusive, to the polar state[2,3]. In 1965, Anderson and Blount proposed a ferroelectric metal state based on the assumption that free electrons do not interact strongly with TO phonons[1]. Ensuing studies raised various possibilities of making ferroelectric polar metals, but the underlying physics has remained unsettled[2,3]. The first polar metal $LiOsO_3$ was identified[12] nearly half a century after the original theoretical prediction, and it was later confirmed by further studies using extreme ultraviolet second harmonic generation excited by an X-ray free electron lasers[13]. Recently, two-dimensional bilayer transition metal dichalcogenides $WTe_2$ and $T_d$-$MoTe_2$ were found to be first ferroelectric metals[14,15]. Some design strategies have been proposed to achieve polar metal states[2,3], such as carrier doping in polar systems[16,17] or constructing artificial two-dimensional systems[18-20]. However, report of realized polar metal state is extremely rare, especially in single-phase bulk materials.

In this work, we devised a new strategy to realize the elusive polar metal state via pressure tuning of the crystal and electronic structures and also generate superconductivity in the polar metal state. Ferroelectricity (FE) exists in metal phosphorous trichalcogenides ($MPX_3$/$M_2P_2X_6$, M'M''$P_2X_6$), and the FE critical temperature depends on specific metal and X atoms and stacking of $MPX_3$ layers. Pressure induces insulator-metal transitions and superconductivity in $MPX_3$ compounds, such as $FePS_3$, $FePSe_3$, $NiPSe_3$ and $Pb_2P_2S_6$ (ref.[21-23]). It is promising to explore polar metal state and superconductivity in this family of compounds under pressure. Recently, ferroelectric $SnPS_3$ (or $Sn_2P_2S_6$) was found to be a superconductor when it transformed under pressure from the original ferroelectric $Pc$ phase to paraelectric $P2_1/c$ and then to a $P2_1$ chiral polar phase[24]. Similar transitions among $Pc$, $P2_1$ and $P2_1/c$ phases were found in artificial polar metal $NdNiO_3$ stabilized at an oxide interface[18]. An earlier study also found possible signatures of a polar metal state in $NiPS_3$ under pressure[25]. Overall, however, there is still a lack of solid evidence of polar metal states in any of the metal phosphorous trichalcogenides. Here, we provide compelling evidence of robust polar metal state at room temperature in $HgPSe_3$ under pressure and, surprisingly, also find that superconductivity emerges in the polar metal state as temperature drops to 11 K, which is much higher than those previously reported in doped $SrTiO_3$ or bilayer $MoTe_2$ (ref.[15,26]).

**Pressure driven structural phase transitions and interlayer bonding changes in $HgPSe_3$**

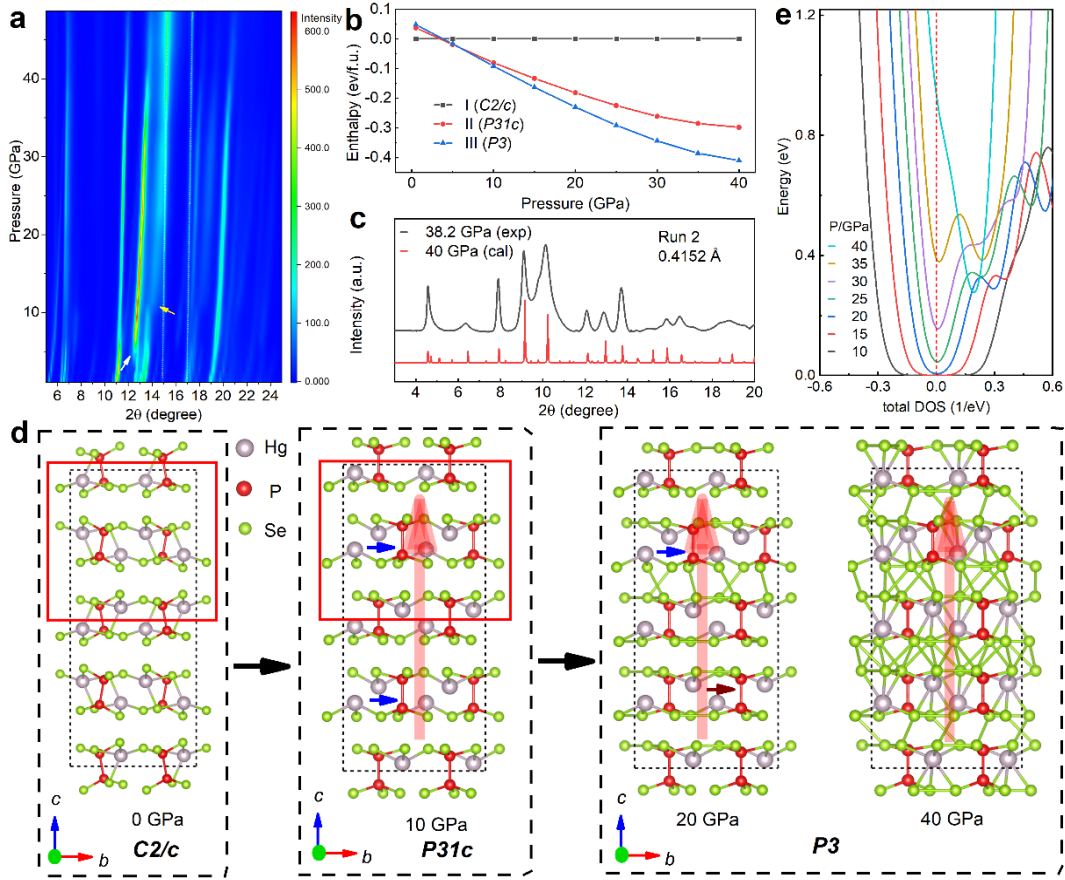

**Fig.1 Pressure induced structural phase transitions of HgPSe$_3$. a,** Integrated X-ray diffraction (XRD) patterns obtained from a 2D panel detector. Two phase transitions (indicated by the arrows) are visible: one near 4.7 GPa where a new peak near 2θ ≈ 12.5° appears, and the other around 11.4 GPa where a peak shift and broadening occurs near 2θ ≈ 14°. **b,** Calculated enthalpy of various phases. At 3.6 to 7 GPa, a polar *P31c* phase is energetically favored. Further compression drives more layer sliding, resulting in a *P3* phase above ~7 GPa. **c,** Comparison of simulated XRD of the *P3* phase and experimental XRD near 40 GPa. **d,** Structural evolution and relation of the *C2/c*, *P31c* and *P3* phases, showing interlayer bonding changes in the *P3* phase at different pressures, where partial interlayer bonding appears at 20 GPa, while much stronger bonding forms at 40 GPa. The vertical red arrows show the direction of electric polarization. The blue and brown arrows indicate the relative position in each phase with 1/6*b* slipping and 1/2*b* slipping of the layer from the original *C2/c* phase, respectively. **e,** Calculated electronic density of states (DOS) of the *P3* phase at various pressures, with metallization occurring at 20 GPa.

We examined structural evolution of HgPSe$_3$ under pressure by powder X-ray diffraction (XRD) supported by theoretical analysis. At ambient pressure, HgPSe$_3$ adopts a monoclinic *C2/c* structure, which is centrosymmetric. A phase transition occurs near 4.7 GPa, as seen in Fig.1a and Extended Data Fig. 1, signaled by increasing peak intensity at 2θ ≈ 12.5°. Another phase transition occurs at 11.4 GPa, as indicated by a sudden shift and broadening of the peak at 2θ ≈ 14°. There is no sign of any further phase transition up to 44.8 GPa, the highest pressure reached.

At ambient conditions, the HgPSe$_3$ crystal comprises stacked quasi-two-dimensional layers that couple via interlayer van der Waals interactions to form the bulk structure. Such layered bulk structures are prone to deformation via interlayer sliding because of the weak interactions between adjacent layers. To probe such scenarios in HgPSe$_3$ and determine its stable interlayer stacking configurations under pressure, we performed structure search and optimization via an automatic relaxation procedure. Comparing calculated results with the experimental data, we determined the structural evolution route in the experimental pressure range. Calculated enthalpy results (Fig.1b) indicate that the low-pressure *C2/c* phase (phase I) becomes unstable above 3.6 GPa and transitions to a new phase in *P31c* space group symmetry (phase II), and the predicted critical pressure is in good agreement with the experimentally determined pressure 4.7 GPa (Fig. 1a). Further compression stabilizes another phase in *P3* space group symmetry (phase III) via more interlayer sliding at 11.4 GPa (Fig. 1a), which is slightly higher than the calculated value (Fig. 1b), indicating possible hinderance by a kinetic barrier for the phase transition. Calculated phonon spectra confirm the stability of the *P31c* phase (phase II, see Extended Data Fig. 2) and *P3* phase (phase III, see Extended Data Fig. 3). It is noted that the *P31c* phase of HgPSe$_3$ shares the same structure with the ferroelectric *P31c* phase in bulk CuInP$_2$Se$_6$ and nanoscale CuInP$_2$S$_6$ (ref.[27,28]). The calculated XRD patterns are well matched with experimental data for the *P3* phase (Fig. 1c) and *P31c* phase (Extended Data Fig. 4). The main difference between the ambient *C2/c* phase and high-pressure *P31c* phase lies in the stacking form of HgPSe$_3$ monolayers and the Se$_3$P-PSe$_3$ units. The *P31c* phase is stabilized via an interlayer sliding by a distance of *b/6* from the *C2/c* phase along the *y* direction, and the tilted Se$_3$P-PSe$_3$ units in the *C2/c* phase are aligned along the c-axis without any tilt in the *P31c* phase. Such a structural transition can be seen clearly in Fig.1d. From the *P31c* to *P3* phase, there is a layer sliding by a distance of *b/2*, as indicated by the brown arrow in Fig.1d. We also examined the evolution of the intralayer and interlayer bonding in the *P31c* and *P3* phases under pressure. At 10 GPa, HgPSe$_3$ still adopts a layered structure with individual Se$_3$P-PSe$_3$ units; at 20 GPa, the Se$_3$P-PSe$_3$ units are connected via partial Se-Se bonding with enhanced intralayer and interlayer interactions; at 40 GPa, these interactions become much stronger, and the crystal shows clear 3D features instead of a 2D layered structure. We calculated electron density of states, and the results (Fig.1e) show a finite band gap at P ≤ 15 GPa, which disappears, signaling an insulator-to-metal transition, at P ≥20 GPa.

**Pressure driven nonpolar-polar transition and metallicity in HgPSe$_3$**

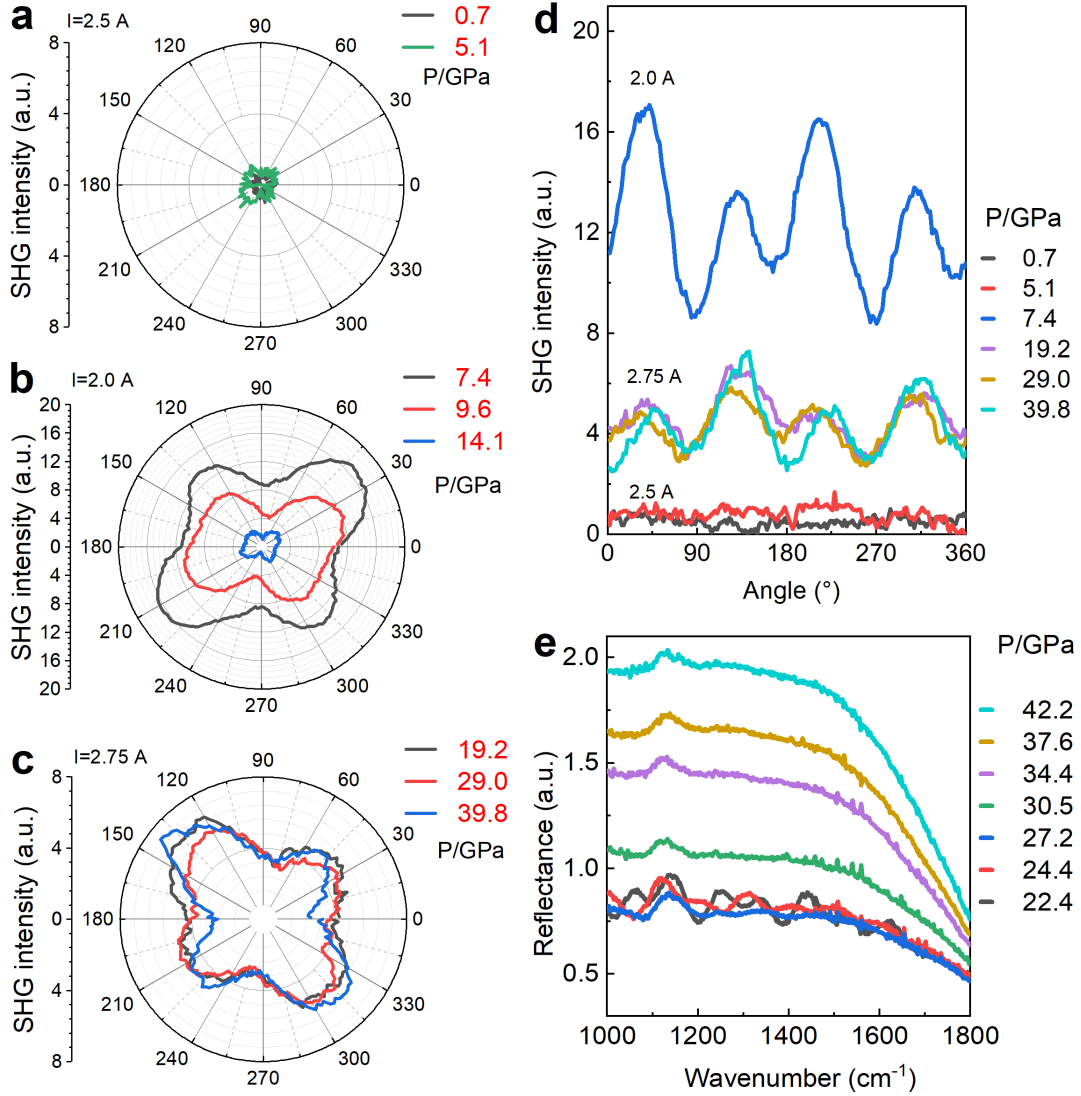

**Fig. 2 Pressure driven nonpolar-polar transition in HgPSe₃ revealed by second harmonic generation (SHG) and metallicity by infrared spectroscopy. a**, SHG signal is absent at 0.7 GPa, which is consistent with the centrosymmetric structure of the pristine *C2/c* phase, but signal appears at 5.1 GPa, indicating that the structure has deviated from the centrosymmetric phase. **b-c**, Strong SHG signal is present at higher pressures, indicative of an emergent polar state. **d**, Linear plot of SHG signals at various pressures. **e**, Metallicity in HgPSe₃ revealed by enhanced reflection of infrared spectroscopy at the low energy region at pressures above 20 GPa.

HgPSe₃ crystallizes in a centrosymmetric *C2/c* structure at the ambient condition. A crystal with centrosymmetric symmetry does not respond to second harmonic generation (SHG), which is confirmed in the low-pressure *C2/c* phase of HgPSe₃ that shows SHG signals fluctuating near zero at 0.7 GPa (Fig. 2a). Upon further compression, SHG signal starts to appear at 5.1 GPa and turns more pronounced at higher pressures (Fig.2b). Nearly symmetric SHG signals were observed at 7.4 GPa, and the intensity declines somewhat at higher compression, but SHG signals can still be observed at 39.8 GPa under strong laser power. The pressure evolution of the SHG intensity is plotted in Fig. 2d, showing the same broad trend and variation pattern in the entire experimental pressure range without any indication of notable changes in the underlying bonding structure. Meanwhile, however, signs of more subtle structural changes are visible, especially the change in SHG intensity from 7.4 GPa to 19.2 GPa, where the orientation of the maximal SHG signal rotates by 90° (Fig.2b-c). We noticed that collecting signals from different positions of the same sample sometimes makes the obtained SHG pattern asymmetric, suggesting that SHG is sensitive to local bonding structures, especially near the sample edges. To check the repeatability of the measured signals, we performed another set of SHG experiments, and the results confirm both the consistent broad trend and pattern of the SHG signals and the sensitivity to pressure driven bonding changes (Extended Data Fig. 5). The symmetric SHG pattern in the *P31c* phase of $HgPSe_3$ is well matched with that found in the ferroelectric *P31c* or *Cc* phase of nanoscale $CuInP_2S_6$ (ref.[28,29]), indicating that the *P31c* phase of $HgPSe_3$ is likely also a room-temperature ferroelectric. In addition, the *P3* phase of $HgPSe_3$ can be regarded as a slight variant of the *P31c* phase, and it is also a polar phase. The polarization in the *P31c* and *P3* phases stems from sliding ferroelectricity in these structures.

The pressure-driven nonpolar-polar transition in $HgPSe_3$ is highly unusual, since polar states tend to be suppressed under compression. For example, the polar *I4mm* phase of SnP transforms to a nonpolar cubic phase under pressure at about 0.4 GPa[30]; the FE state in $NbOX_2$ (X = Cl, I) is suppressed by pressure at about 6 GPa[31]. Although $CuInP_2S_6$ shows some anomaly with enhanced FE state at low pressure (~0.26 GPa), its FE state is also suppressed by pressure above 2 GPa[32]. In sharp contrasty, pressure induces and maintains an emergent robust polar state in $HgPSe_3$, which offers a unique platform to explore novel quantum orders and phase transitions.

Calculated electronic density of states revealed a metallic state in $HgPSe_3$ under pressures above 20 GPa (Fig. 1e), suggesting a rare polar metal state in this compound. To verify this prediction, we conducted reflectance measurements employing infrared spectroscopy techniques. The results (Fig. 2e) show that the reflectance of compressed $HgPSe_3$ sample increases suddenly above 27.2 GPa without any interference (periodically fluctuating) signals, and the signal intensity exhibits a monotonously increasing trend with rising pressure. Such a change reflectance is a clear sign of a metallization process. Between 22.4 and 27.2 GPa, some interference signals are visible, which is understandable since the sample just turned into a metallic state above 20 GPa and the electronic density of states is still low. Consequently, $HgPSe_3$ is in a semimetal state that allows transmission of infrared light, resulting in the interference in spectroscopy measurements. The interference weakens with suppressed fluctuation amplitudes and longer fluctuation periods as pressure rises from 22.4 to 27.2 GPa, reflecting enhanced metallicity of the sample.

**Pressure driven Insulator-metal transition and superconductivity in $HgPSe_3$**

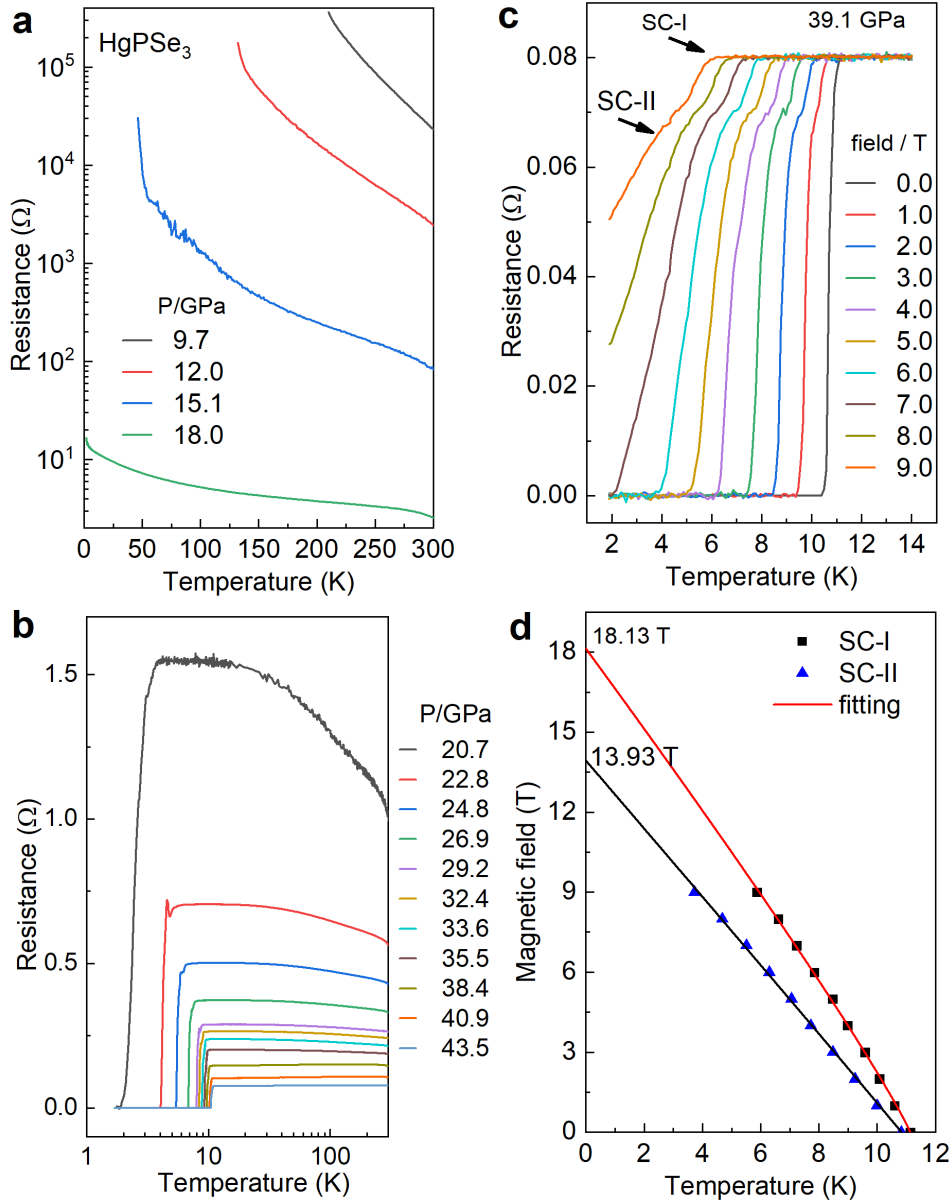

**Fig.3 Transport properties of HgPSe$_3$ under pressure. a**, Insulating/semiconducting behaviors at low pressures (9.7-18.0 GPa). Below 9.7 GPa, the resistance is out of the equipment detection limit. **b**, Superconducting transition under high pressures (20.7-43.5 GPa). The sample becomes more metallic and $T_c$ increases monotonously with rising pressure. **c**, Superconducting transitions under various magnetic fields at 39.1 GPa. A double superconducting (SC) transition behavior is observed as marked by the arrows. **d**, The $H$-$T_c$ relation and empirical fitting using the formula $H_{c2}(T) = H_{c2}*(1-T/T_c)^{1+a}$ for the SC phases.

To measure transport properties, we exfoliated a thin HgPSe$_3$ flake with a thickness of ~5-10 μm from a bulk crystal and cut it into a well-defined rectangular shape for high-pressure resistance measurements. Temperature dependent transport measurements were conducted using a standard four-probe geometry. At ambient conditions, the resistance of the HgPSe$_3$ sample was too large to be measured by a portable multimeter. To ensure data quality, we carried out low-temperature transport measurements starting at ~9.7 GPa when the multimeter showed a resistance on the scale of 10-100 KΩ. Further compression enhanced the conductivity of the sample, and the measured resistance data (Fig.3a) indicate that the sample is still highly insulating at low temperatures with the resistance exceeding the upper limit of our instrument range at pressures of ~9.7 GPa. When the pressure is increased to ~18.0 GPa, a full set of R-T data was obtained from 1.7 K to 300 K. The resistance at 18.0 GPa is 2.5 Ω at 300 K and 16.1 Ω at 1.8 K, suggesting the sample to be in a semimetal state or near the boundary of a semimetal state[33-35]. Surprisingly, we found $HgPSe_3$ entering a superconducting state at ~20.7 GPa with onset and zero-resistance critical temperatures of $T_c^{onset} \approx 3.1$ K and $T_c^{zero} \approx 1.8$ K, respectively, as shown in Fig.3b. Further compression elevates $T_c^{onset}$ to ~11.0 K and $T_c^{zero}$ to ~10.4 K at 43.5 GPa, which is the highest pressure measured, with a very sharp SC transition. While the rate of increasing $T_c$ slows down with rising pressure, the trend is not saturated at 43.5 GPa, suggesting higher $T_c$ under further compression. The R-T curves above the SC transition temperature shows a semimetal behavior with weak temperature dependent resistance[33-35], which further confirms the theoretically predicted and infrared measured metallic behavior. Similar semimetal behaviors were observed in other layered bulk compounds under pressure[24,36].

Superconductivity in $HgPSe_3$ is further confirmed by measuring the effect of magnetic field on $T_c$. The results (Fig.3c) clearly show that magnetic field gradually suppresses superconductivity as $T_c$ shifts to lower temperature and the transition width becomes broader with increasing magnetic fields. Interestingly, the resistance curves exhibit features indicating two superconducting transitions (SC-I and SC-II), which become clearer at higher magnetic fields. This phenomenon may stem from the presence of deviatoric stress at high pressures, and this result suggests that the properties of the layered $HgPSe_3$ crystal are sensitive to pressure environment. We also performed an empirical fitting to extract the upper critical field at zero temperature using the formula $H_{c2}(T) = H_{c2}^{*}(1-T/T_c)^{1+a}$. The fitting results (Fig.3d) give $\mu_0 H_{c2} = 18.1$ T for the SC-I phase, which is close to but lower than the Bardeen-Copper-Schrieffer (BCS) weak-coupling Pauli paramagnetic limit of $\mu_0 H_p = 1.84 T_c = 20.2$ T for $T_c = 11.0$ K, suggesting an absence of Pauli pair breaking[37]. A slightly lower $\mu_0 H_{c2} = 13.9$ T is obtained for the SC-II phase.

An intriguing question is whether superconductivity in $HgPSe_3$ emerges from the polar metal state. To probe this issue, we conducted low-temperature XRD measurements in the pressure range of 44-48 GPa. To make sure that the pressure determination is accurate, a calibration using a Pt inner pressure marker was carried out. There is some pressure change during the cooling process, due to the thermal expansion of the BeCu sample cell. Overall, the pressure change is relatively small. The XRD results (Extended Data Fig. 6) at different temperatures are almost identical to that at 38.2 GPa (Fig.1c). There is no signature of any structural phase transition, just a stable *P3* phase, in the examined pressure range. The lowest temperature for XRD measurements is 9 K, which is below the superconducting transition temperature (since $T_c^{zero}$ is higher than 10 K above 39.1 GPa, as seen in Figs. 3b and 3c). These results show that superconductivity in $HgPSe_3$ indeed emerges from and coexist with the polar metallic *P3* phase at low temperatures.

**The property-pressure phase diagram of $HgPSe_3$**

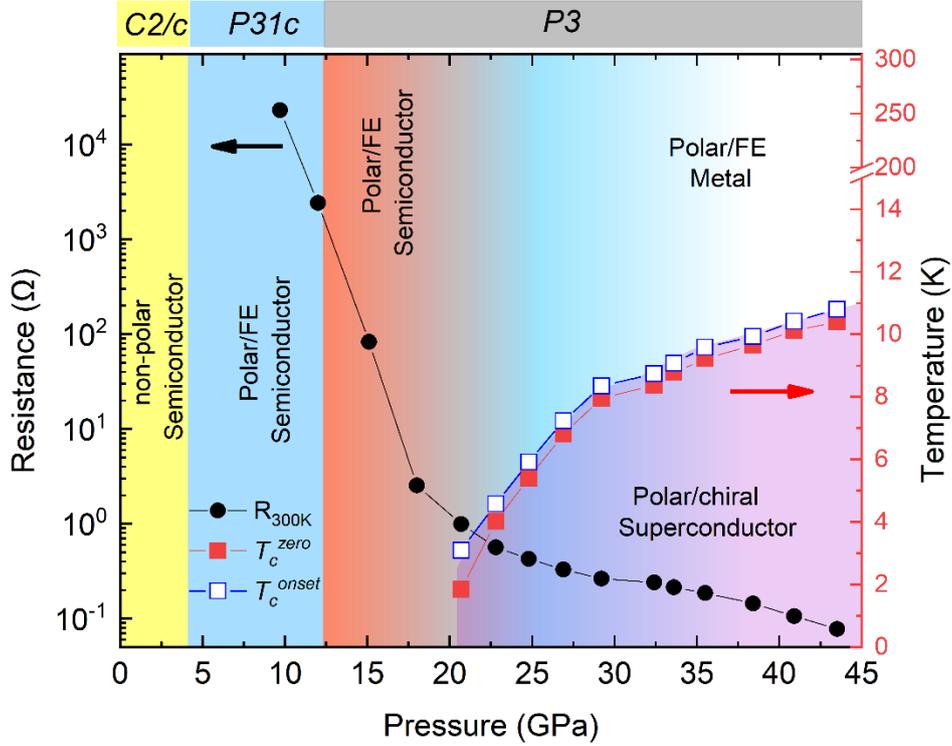

**Fig. 4 The property-pressure phase diagram of $HgPSe_3$.** $HgPSe_3$ undergoes a structural phase transition near 3.6 GPa, which is a nonpolar-polar transition with significant SHG response in the higher-pressure *P31c* phase. At 12 GPa, the *P31c* phase transforms to a chiral and polar *P3* phase. An insulator-metal transition occurs near 20 GPa, and superconductivity emerges afterwards with $T_c$ showing a positive pressure dependence and the highest $T_c$ reaches ~11.0 K at 43.5 GPa. The polar metal state is robust at room temperature and coexists with the superconducting state at lower temperatures in a wide pressure range.

The results from our joint experimental and computational study give a systematic description of crystal stability and evolution, SHG, infrared spectroscopy and electrical transport measurements of $HgPSe_3$ in a wide range of pressure and temperature, offering a clear and rich picture about this rare bulk polar metal that also hosts superconductivity above 20 GPa with $T_c$ up to 11.0 K. A comprehensive phase diagram summarizing the key properties of $HgPSe_3$ is shown in Fig. 4. It is seen that below 3.6 GPa, $HgPSe_3$ is a nonpolar semiconductor in *C2/c* space group symmetry, then transforms to a polar *P31c* phase, signaled by strong SHG responses. Further compression drives the *P31c* phase into a polar/chiral *P3* phase, where the polarity is reduced, with weaker SHG response than that in the *P31c* phase. Structurally, the *P31c* phase stems from an interlayer sliding with a lattice shift of a distance of 0.1667*b (=b/6) based on the *C2/c* phase, while an extra partial interlayer sliding by a distance of 0.5*b occurs in the *P3* phase. Therefore, the polar states in the *P31c* and *P3* phases are rooted in sliding ferroelectricity[38], which has been observed in various layered systems[15,39,40]. Post structural phase transitions, the *P31c* and *P3* phases of HgPSe$_3$ are in insulating states until further compression drives an insulator-metal transition in the *P3* phase near 20 GPa, and superconductivity emerges and becomes more pronounced with rising pressure. The highest $T_c$ reaches ~11.0 K, which is much higher than in doped SrTiO$_3$ ($T_c^{max} \approx$ 0.6-0.7 K in ref.[26]) and bilayer MoTe$_2$ ($T_c^{max} \approx$ 2.5 K in ref.[15]), setting a record for superconductivity in polar metals. Low-temperature XRD spectra taken at high pressure show no phase transition in the *P3* phase during the superconducting transition, indicating that compressed HgPSe$_3$ is a superconductor stemming from the polar/FE metal ground state. Further work is required to probe the potential FE metal state in the *P3* phase of HgPSe$_3$ which, if realized, would be an extremely rare case of ferroelectric and chiral superconductor, providing an ideal platform to study pertinent FE quantum critical point and its correlation with superconductivity.

**Discussion**

Distinct from previously proposed strategies to realize the polar metal state[2,3], the pressure-driven polar metal state devised in this work is achieved via a two-steps process. First, an interlayer sliding mode breaks the centrosymmertry of the crystal structure in the *C2/c* symmetry, turning the originally nonpolar state into a polar one in a non-metallic state with a sizable electronic band gap. Then, further compression shortens the interlayer bonds to close the band gap and then enhance conductivity, leading to the metallic state. The resulting room-temperature polar metal state exists in a bulk form rather than two-dimensional systems as in the previously reported bilayer systems or artificial oxide interfaces[2,3]. Bulk polar metals are extremely rare, let alone one that is robust at room temperature as found in HgPSe$_3$. Here, the polar state is hosted by a chiral *P3* phase, which is a variant of a typical ferroelectric *P31c* phase; further work may achieve switchable FE states and find correlation between FE and superconductivity. Our discovery of interlayer sliding induced polar metal state in the bulk form suggests that the vast families of layered materials may offer highly promising and productive platforms for searching and finding more polar metals via pressure or strain engineering. The concept of sliding ferroelectricity[38] may be extended to achieve room-temperature polar metal state using metallic or even superconducting monolayers, using, e.g., NbSe$_2$ (ref.[41]) and Bi$_2$Sr$_2$CaCu$_2$O$_{8+\delta}$ (ref.[42]), as the building blocks. The resulting combination of artificially constructed FE and SC states or other exotic quantum states would greatly enrich the scope of polar metals and bring about distinct novel physics.

**Conclusions**

The robust polar metal state discovered in the layered bulk HgPSe$_3$ crystal at room temperature introduces a new paradigm for finding and studying this exotic quantum state of matter. Our work also opens an intriguing avenue to probe novel physics coexisting in the polar metal state, exemplified by the superconductivity found in HgPSe$_3$ that exhibits record high $T_c^{SC}$ compared to recently reported artificially constructed polar metals. The polar *P31c* and *P3* phases of HgPSe$_3$ are rooted in interlayer sliding from the original nonpolar *C2/c* phase; the nonpolar-polar transition occurs at relatively low pressures around 3.6 GPa, followed by an insulator-metal transition near 20 GPa, where superconductivity emerges at low temperatures. The SC critical temperature exhibits a monotonous increasing trend with pressure, from ~3.1 K ($T_c^{onset}$) at 20.7 GPa to ~11.0 K ($T_c^{onset}$) at 43.5 GPa. Low-temperature high-pressure XRD measurements find no phase transition during the superconducting transition, indicating that superconductivity emerges in the polar metal ground state. The large families of $MPX_3$, $M_2P_2X_6$ and $M'M''P_2X_6$ compounds possess versatile interlayer stacking modes and, consequently, may be able to adopt diverse structural phases stabilized by pressure or strain conditions with novel properties stemming from the interlayer sliding mechanism. Work along this line may prove fruitful for discovery of exotic physic such as magnetic polar metal state and its coexistence with superconductivity and topological order.

**Methods**

*In situ* **synchrotron x-ray diffraction under pressure**

A symmetric DAC with Boehler-type diamonds (opening angle > 60°) and WC seats was used for the in situ XRD experiments at room temperature. A pre-compressed rhenium gasket was drilled with a 120 μm hole to serve as a sample chamber. The powder sample was finely grounded and loaded into the sample chamber. Silicon oil was used as the pressure transmitting medium. The XRD patterns were collected at the Beamline 4W2 with a wavelength of 0.6199 Å, a beamsize of ~33*12 microns and a Pilatus 2D detector at the Beijing Synchrotron Radiation Facility. $CeO_2$ was used to calibrate the instrument parameters including sample to detector distance, beam center and detector tilt. A gas membrane system was employed to add the pressure *in situ,* and the pressure was calibrated by an online ruby system. The 2D diffraction images were then processed by the Dioptas software to obtain the integrated linear plot of XRD patterns.

Another high-pressure and low-temperature XRD experiment was carried out at the BL10XU beamline of Super Photon ring-8 (SPring-8, Japan) with a wavelength of λ = 0.4152 Å. A custom-designed low temperature BeCu cell was utilized with two opposing 300-μm culet Boehler-type diamond anvils with ~60 degree open angles. The data was collected by a Perkin Elmer digital x-ray flat panel detector, and $CeO_2$ was used to calibrate the instrument parameters including sample to detector distance, beam center and detector tilt. To achieve a lower temperature, the low temperature sample chamber was evacuated by a pumping system with both a mechanical pump and a turbo molecular pump for more than 4 hours. In high-pressure XRD experiments at low temperature, we employed a Pt inner pressure marker to calibrate the pressure. Meanwhile, to reduce possible pressure change due to the thermal expansion of cells, the pressure was pre-locked by a set of pressure-locking screws.

**Second harmonic generation detection under pressure**

Another symmetric DAC with 300 μm culet diamond (Type-II) was used for the SHG experiments.

A rhenium gasket was pre-compressed and drilled with a 120 μm hole as a sample chamber. KBr was used as the pressure transmitting medium for SHG experiments, conducted in a home-designed optical system (Ideaoptics, China). The exciting light source is a 1,064 nm fiber laser. SHG signal was collected by a photomultiplier tube (Thorlabs Inc., PMT1000) by rotating the polarizer.

**High-pressure temperature dependent electrical transport measurements**

Four-probe resistance measurements were performed to investigate the electric transport properties of $HgPSe_3$ under high pressure using a BeCu-type diamond anvil cell with 300 μm culet diamonds. A rhenium gasket with a c-BN+epoxy mixture insulating layer was drilled with a 100 μm hole to serve as a sample chamber. KBr was used as the pressure transmitting medium. The R-T curves were collected up to 43.5 GPa in a commercial cryostat from 1.7 K to 300 K by a Keithley 6221 current source and a 2182A nanovoltmeter. The R-T curves near the superconducting transition region were collected in magnetic fields up to 9.0 T. A thick flake was exfoliated from bulk $HgPSe_3$ crystal and cut into a long-strip shape before being loaded into the sample chamber. Four Pt electrodes were contacted to the sample in the standard four-probe configuration. The pressure of the sample chamber was calibrated by ruby fluorescence before and after cooling-warming cycling.

**Theoretical calculations**

Our density functional theory (DFT) calculations were performed using the Vienna *ab initio* simulation package[43] with the all-electron projector augmented wave method[44]. The generalized gradient approximation (GGA) revised for solids (PBEsol)[45] was used as the exchange-correlation functional. The valence states $5d^{10}6s^2$ for Hg, $3s^23p^3$ for P, and $4s^24p^4$ for Se were used with the energy cutoff of 500 eV for the plane wave basis set. The Brillouin-zone sampling was performed using a 6×3×3 *k*-point grid. The structures were fully relaxed until the total energy difference is less than $1×10^{-6}$ eV and convergence criteria for atomic forces was set to be $10^{-2}$ eV/Å. The high-pressure phases are explored up to 50 GPa *via* a one-layer by one-layer slip mechanism[46] along the in-plane *a* or *b* directions under pressure. The electronic density of states were calculated using the modified Becke-Johnson (mBJ) functional[47]. Phonon spectra and density of states were calculated using the MedeA-Phonon code[48].

**Data availability**

All data that support the findings of this study are available from the corresponding author upon reasonable request.

**Acknowledgments**

This work was supported by the National Key R&D Program of China (Grants No. 2021YFA1400300 and 2020YFA0711502), and the Major Program of National Natural Science

Foundation of China (22090041) and National Natural Science Foundation of China (Grant No. 12374050, 12004014, U1930401, No. 92263202, and No. 12374020) and the Strategic Priority Research Program of the Chinese Academy of Sciences (Grant No. XDB33000000). Part of the experimental work was carried out at the Synergic Extreme Condition User Facility.

**Author contributions:** F. H. and B.B.Y. conceived the work; J.T.W. did the theoretical calculation; B.B.Y provided the single crystal; W.Z., B.B.Y., F.H. performed the synchrotron XRD, SHG, and Infrared experiments; F.H. loaded the DAC for transport measurements, and X.H.Y. collected the electrical transport data; F.H., B.B.Y., J.T.W. analyzed the results; X.L.W., Z.X.C., and H.K.M. interpreted the data. F.H. and C.C. wrote the paper with comments from all authors.

**Competing interests:** The authors declare no competing interests.

**References**

1  Anderson, P. W. & Blount, E. I. Symmetry Considerations on Martensitic Transformations: "Ferroelectric" Metals? *Phys. Rev. Lett.* **14**, 217-219, (1965).

2  Zhou, W. & Ariando, A. Review on ferroelectric/polar metals. *Jpn. J. Appl. Phys* **59**, SI0802, (2020).

3  Bhowal, S. & Spaldin, N. A. Polar metals: principles and prospects. *Annu. Rev. Mater. Res.* **53**, 53-79, (2023).

4  Eerenstein, W., Mathur, N. & Scott, J. F. Multiferroic and magnetoelectric materials. *Nature* **442**, 759-765, (2006).

5  Fiebig, M., Lottermoser, T., Meier, D. & Trassin, M. The evolution of multiferroics. *Nat. Rev. Mater.* **1**, 1-14, (2016).

6  Ginley, D. S. & Bright, C. Transparent conducting oxides. *MRS bulletin* **25**, 15-18, (2000).

7  Stewart, G. Superconductivity in iron compounds. *Rev. Mod. Phys.* **83**, 1589, (2011).

8  Moore, E. A. & Smart, L. E. *Superconductivity* (*Solid State Chemistry*) 347-362 (CRC Press, 2020).

9  Nomura, Y. & Arita, R. Superconductivity in infinite-layer nickelates. *Rep. Prog. Phys.* **85**, 052501, (2022).

10  Chiu, C.-K., Teo, J. C., Schnyder, A. P. & Ryu, S. Classification of topological quantum matter with symmetries. *Rev. Mod. Phys.* **88**, 035005, (2016).

11  Lee, S. *et al.* Anomalously low electronic thermal conductivity in metallic vanadium dioxide. *Science* **355**, 371-374, (2017).

12  Shi, Y. *et al.* A ferroelectric-like structural transition in a metal. *Nature Mater.* **12**, 1024-1027, (2013).

13  Berger, E. *et al.* Extreme ultraviolet second harmonic generation spectroscopy in a polar metal. *Nano letters* **21**, 6095-6101, (2021).

14  Fei, Z. *et al.* Ferroelectric switching of a two-dimensional metal. *Nature* **560**, 336-339, (2018).

15  Jindal, A. *et al.* Coupled ferroelectricity and superconductivity in bilayer $T_d$-MoTe$_2$. *Nature* **613**, 48-52, (2023).

16  Kolodiazhnyi, T., Tachibana, M., Kawaji, H., Hwang, J. & Takayama-Muromachi, E. Persistence of ferroelectricity in BaTiO$_3$ through the insulator-metal transition. *Phys. Rev.*

| | |
|---|---|
| | *Lett.* **104**, 147602, (2010). |
| 17 | Gu, J.-x. *et al.* Coexistence of polar distortion and metallicity in PbTi$_{1-x}$Nb$_x$O$_3$. *Phys. Rev. B* **96**, 165206, (2017). |
| 18 | Kim, T. *et al.* Polar metals by geometric design. *Nature* **533**, 68-72, (2016). |
| 19 | Ma, C. & Jin, K. Design strategy for ferroelectric-based polar metals with dimensionality-tunable electronic states. *Sci. China-Phys. Mech. Astron.* **61**, 1-6, (2018). |
| 20 | Zhou, W. X. *et al.* Artificial two-dimensional polar metal by charge transfer to a ferroelectric insulator. *Commun. Phys.* **2**, 125, (2019). |
| 21 | Wang, Y. *et al.* Emergent superconductivity in an iron-based honeycomb lattice initiated by pressure-driven spin-crossover. *Nature Commun.* **9**, 1-7, (2018). |
| 22 | Sun, H. *et al.* Coexistence of zigzag antiferromagnetic order and superconductivity in compressed NiPSe$_3$. *Mater. Today Phys.* **36**, 101188, (2023). |
| 23 | He, X. *et al.* Pressure-induced superconductivity in the metal thiophosphate Pb$_2$P$_2$S$_6$. *Phys. Rev. Mater.* **7**, 054801, (2023). |
| 24 | Yue, B. *et al.* Pressure-induced ferroelectric-to-superconductor transition in SnPS$_3$. *Phys. Rev. B* **107**, L140501, (2023). |
| 25 | Harms, N. C. *et al.* Symmetry progression and possible polar metallicity in NiPS$_3$ under pressure. *NPJ 2D Mater. Appl.* **6**, 40, (2022). |
| 26 | Hameed, S. *et al.* Enhanced superconductivity and ferroelectric quantum criticality in plastically deformed strontium titanate. *Nature Mater.* **21**, 54-61, (2022). |
| 27 | Dziaugys, A. *et al.* Piezoelectric domain walls in van der Waals antiferroelectric CuInP$_2$Se$_6$. *Nature Commun.* **11**, 3623, (2020). |
| 28 | Deng, J. *et al.* Thickness-dependent in-plane polarization and structural phase transition in van der Waals ferroelectric CuInP$_2$S$_6$. *Small* **16**, 1904529, (2020). |
| 29 | Liu, F. *et al.* Room-temperature ferroelectricity in CuInP$_2$S$_6$ ultrathin flakes. *Nature Commun.* **7**, 12357, (2016). |
| 30 | Kamitani, M. *et al.* Superconductivity at the polar-nonpolar phase boundary of SnP with an unusual valence state. *Phys. Rev. Lett.* **119**, 207001, (2017). |
| 31 | Ye, L. *et al.* Manipulation of nonlinear optical responses in layered ferroelectric niobium oxide dihalides. *Nature Commun.* **14**, 5911, (2023). |
| 32 | Yao, X. *et al.* Anomalous polarization enhancement in a van der Waals ferroelectric material under pressure. *Nature Commun.* **14**, 4301, (2023). |
| 33 | Mu, Q.-G. *et al.* Suppression of axionic charge density wave and onset of superconductivity in the chiral Weyl semimetal Ta$_2$Se$_8$I. *Phys. Rev. Mater.* **5**, 084201, (2021). |
| 34 | Pavlosiuk, O., Kaczorowski, D., Fabreges, X., Gukasov, A. & Wiśniewski, P. Antiferromagnetism and superconductivity in the half-Heusler semimetal HoPdBi. *Sci. Rep.* **6**, 18797, (2016). |
| 35 | Hirai, D., Matsuno, J., Nishio-Hamane, D. & Takagi, H. Semimetallic transport properties of epitaxially stabilized perovskite CaIrO$_3$ films. *Appl. Phys. Lett.* **107**, 012104, (2015). |
| 36 | Yue, B. *et al.* Insulator-to-Superconductor Transition in Quasi-One-Dimensional HfS$_3$ under Pressure. *J. Am. Chem. Soc.* **145**, 1301-1309, (2023). |
| 37 | Clogston, A. M. Upper Limit for the Critical Field in Hard Superconductors. *Phys. Rev. Lett.* **9**, 266-267, (1962). |

38	Wu, M. & Li, J. Sliding ferroelectricity in 2D van der Waals materials: Related physics and future opportunities. *Proc. Natl. Acad. Sci.* **118**, e2115703118, (2021).

39	Yang, Q., Wu, M. & Li, J. Origin of two-dimensional vertical ferroelectricity in $WTe_2$ bilayer and multilayer. *J Phys. Chem. Lett.* **9**, 7160-7164, (2018).

40	Liu, X. *et al.* Vertical ferroelectric switching by in-plane sliding of two-dimensional bilayer $WTe_2$. *Nanoscale* **11**, 18575-18581, (2019).

41	Xing, Y. *et al.* Ising superconductivity and quantum phase transition in macro-size monolayer $NbSe_2$. *Nano letters* **17**, 6802-6807, (2017).

42	Yu, Y. *et al.* High-temperature superconductivity in monolayer $Bi_2Sr_2CaCu_2O_{8+\delta}$. *Nature* **575**, 156-163, (2019).

43	Kresse, G. & Furthmuller, J. Efficient iterative schemes for ab initio total-energy calculations using a plane-wave basis set. *Phys. Rev. B* **54**, 11169-11186, (1996).

44	Blöchl, P. E. Projector augmented-wave method. *Phys. Rev. B* **50**, 17953, (1994).

45	Perdew, J. P. *et al.* Restoring the density-gradient expansion for exchange in solids and surfaces. *Phys. Rev. Lett.* **100**, 136406, (2008).

46	Wang, J.-T., Chen, C. & Kawazoe, Y. Low-temperature phase transformation from graphite to *$sp^3$* orthorhombic carbon. *Phys. Rev. Lett.* **106**, 075501, (2011).

47	Becke, A. D. & Johnson, E. R. A simple effective potential for exchange. *The Journal of chemical physics* **124**, 221101, (2006).

48	Parlinski, K., Li, Z. & Kawazoe, Y. First-principles determination of the soft mode in cubic $ZrO_2$. *Phys. Rev. Lett.* **78**, 4063, (1997).

**Extended Data**

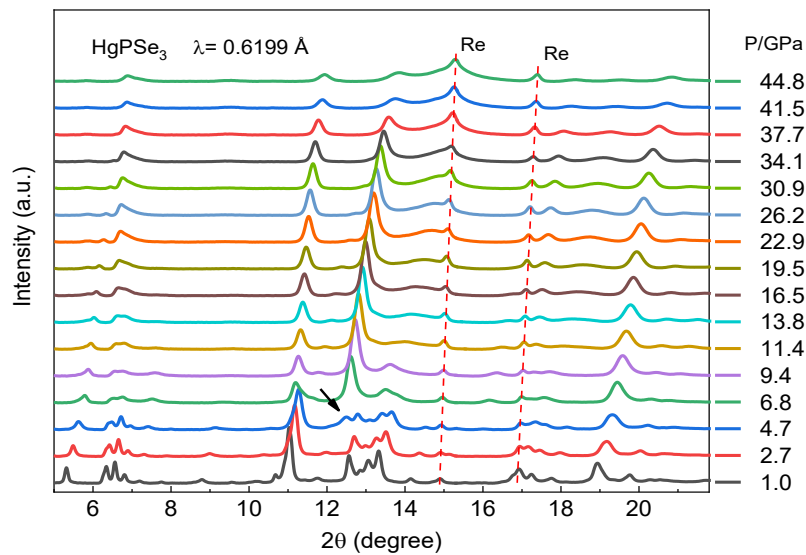

**Extended Data Fig. 1 The linear XRD patterns of HgPSe$_3$ under pressure.** The first structure phase transition was observed above 4.7 GPa, while the second one starts near 11.4 GPa. The diffraction signals from Re gasket are marked by the two dash lines, which display a different pressure dependence than the peaks from HgPSe$_3$.

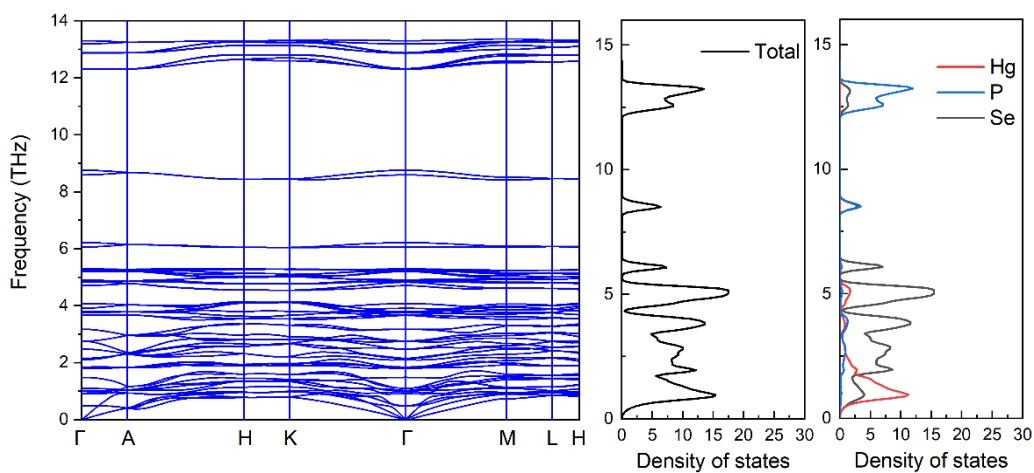

**Extended Data Fig. 2: The phonon spectra and density of states of the *P31c* phase (Phase II) of HgPSe$_3$ at 5 GPa.** The phonon spectra show no imaginary frequencies throughout the entire Brillouin zone, indicating the *P31c* phase is dynamically stable.

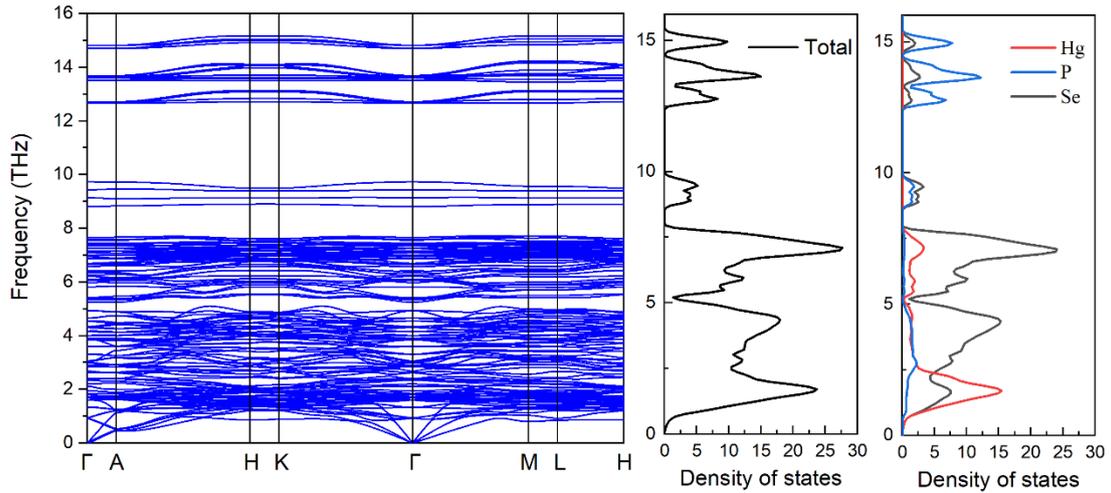

**Extended Data Fig. 3: The phonon spectra and density of states of the *P3* phase (Phase III) of HgPSe₃ at 40 GPa.** The phonon spectra show no imaginary frequencies throughout the entire Brillouin zone, indicating *P3* phase is dynamically stable. The partial density of states from individual atoms is also calculated. The superconductivity is mainly mediated by the low-frequency phonons of Hg atoms and Se atoms, and the high frequency phonons of P atoms.

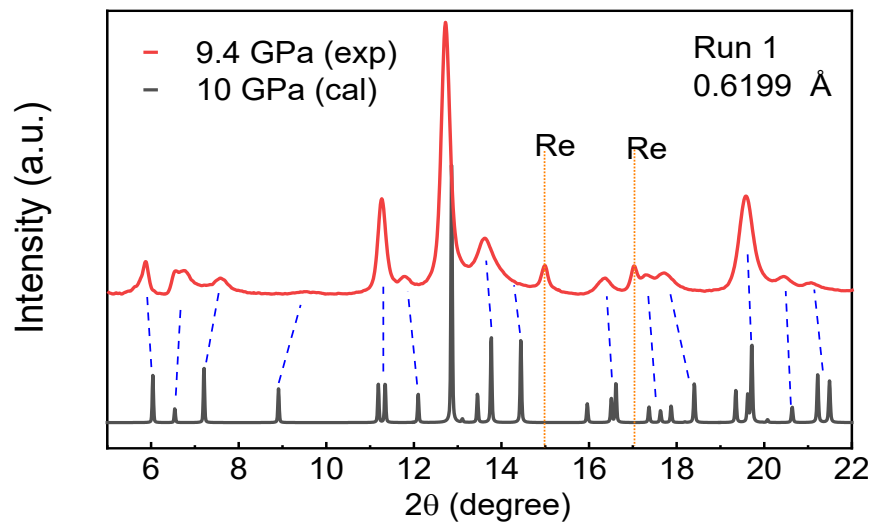

**Extended Data Fig. 4: The experimental and calculated XRD of the *P31c* phase near 10 GPa.** The big difference between the *P31c* phase and original *C2/c* phase is the new intense peak near 2θ =13° marked by the arrow in Fig.1a. The calculated XRD produces such an intense peak very well. There are some deviations between the experimental data and calculated results on XRD patterns, which can be attributed to the difference in lattice parameters, since computation relaxes the lattice automatically to reach the lowest energy of the system, which differs from the experimental values.

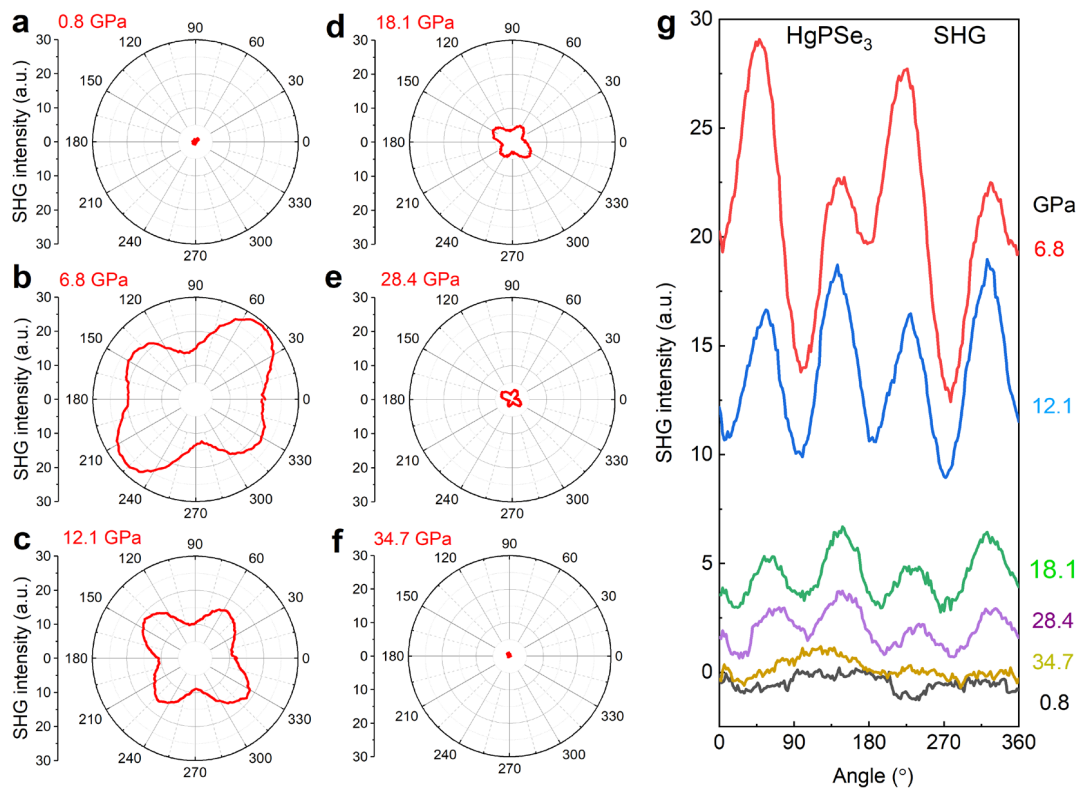

**Extended Data Fig. 5: Another run of the second harmonic generation on HgPSe$_3$.** All SHG signals are plotted in the same scale for easy comparison. **a-f**. SHG signals collected at various pressures, **g.** a linear plot of SHG signals from **a-f**. Overall, when the sample becomes SHG active, its SHG intensity declines with pressure. No clear symmetry change is observed.

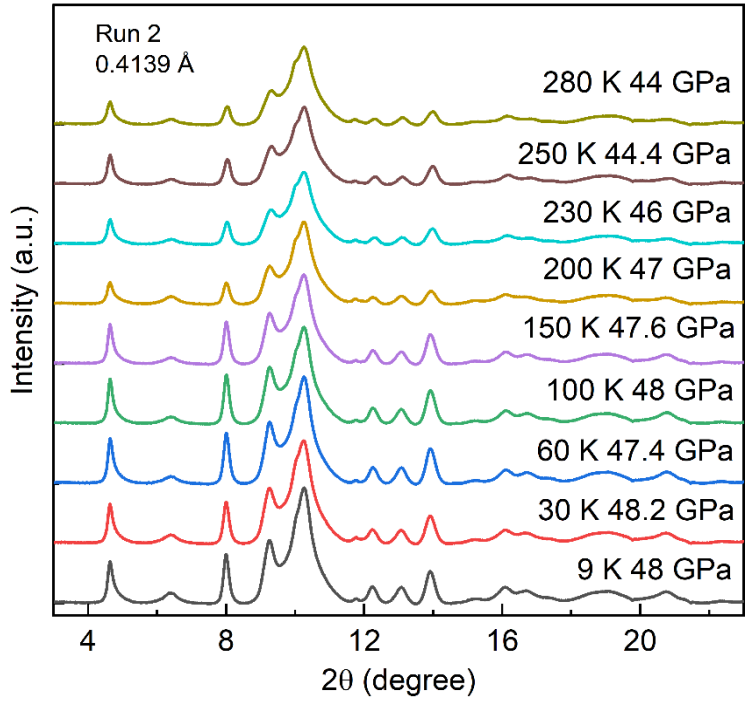

**Extended Data Fig. 6: Low-temperature X-ray diffraction of HgPSe$_3$ near 48 GPa.** The pressure was calibrated by a Pt inner pressure maker in situ. The lowest temperature is at 9 K, which is below the superconducting transition point. No phase transition is observed, confirming a stable *P3* phase as the polar metal ground state.